# Observation of transformation of chemical elements during electric discharge

Urutskoev L.I., Liksonov V.I.

"RECOM" RRC "Kurchatov Institute"

Moscow, Shchukinskaya st. 12-1

tel. 196-90-90, fax 196-1635

e-mail: sergeysmr@mail.ru

Tsinoev V.G.

RRC "Kurchatov Institute"

123182 Moscow, Kurchatov square, 1, tel. 196-73-65

Abstract.

Results of experimental studies of electric explosion, in water, of foils made of extremely pure materials are presented. New chemical elements detected both by spectroscopic measurements during the electric discharge and by a mass-spectrometer analysis of sediments after the discharge have been found to appear. A "strange" radiation associated with the transformation of chemical elements has been registered. A hypothesis has been put forward that particles of the "strange" radiation have magnetic charge.

**Introduction**

The physics of electric explosion of wires in water has been discussed in many papers, reviews, and monographs [1-3], which is mainly due to the large interest to this phenomenon for numerous practical applications. One of such tasks is shattering of concrete foundations. As a rule, devices for this purpose use relatively low-voltage high-capacity batteries (U~ 5 kV) to obtain the required energy storage of order of several tens of kJ. The characteristic discharge time for such batteries is about several hundreds microseconds. Ordinarily, the electric discharges are produced in a narrow (d~20 mm) pit filled with a liquid and the discharge is initiated by exploding wires.

A feature of such an electric explosion scheme is that reflected waves act on the plasma channel produced in a closed volume. They rapidly brake the motion of the channel boundaries, the channel stops expanding, and the pressure at the surface significantly increases. In this processes pressure in the channel can exceed the one attainable at a shock front.

An additional increase of the energy input into the channel can be attained by initiating the discharge by wires made of materials which has a larger thermal effect in reactions with water. These are titanium, zirconium, and beryllium. This possibility has been studied in both the very early [1-4] and later papers [5].

The present work was initially dedicated to study of the efficiency of electric explosion of titanium foils in water to shatter concrete. The experiments revealed that the concrete has been smashed by electric explosion and its fragments have flown away with substantial velocities. A rough estimate of kinetic energy of the fragments based on a visual rapid taking pictures (with a rate of 300 frames per second) was ~8 kJ. It is the willing to study such an effective mechanism of the capacitor battery energy transformation into the kinetic energy of concrete splinters that initiated the experiments results of which are presented in this paper.

**The experiment scheme, diagnostics, and results.**

The scheme of the experiment is shown in Fig. 1. The capacitor battery has been discharged into a foil in water. The energy storage of the battery at a charging voltage of U ~ 4.8 kV was W~50 kJ. The capacitor spark-gaps [6] provided the commutation of the capacitor bank. The energy was supplied to the load by cables 3 with an induction of L=0.4 mcH. The load was a Ti foil that was welded to Ti electrodes 5 by pressure contact welding. The electrodes were mounted on a polyethylene cover 6 which in turn was attached through seals 7 to the explosion chamber 8 also made of polyethylene. The explosion chamber represented a torus which was filled with a liquid through eight holes 9 uniformly drilled around a circle. Distilled water was used as a working liquid in most experiments. The number of load varied from one to eight in different experiments. Analog oscillographs and rapid analog-cipher transformers attached to computers were used to register electric signals. The typical oscillograms of the current and voltage are shown in Fig. 2. Since various diagnostics were used in the experiments, we believe worthwhile describing diagnostics and methods in line with experimental material presentation.

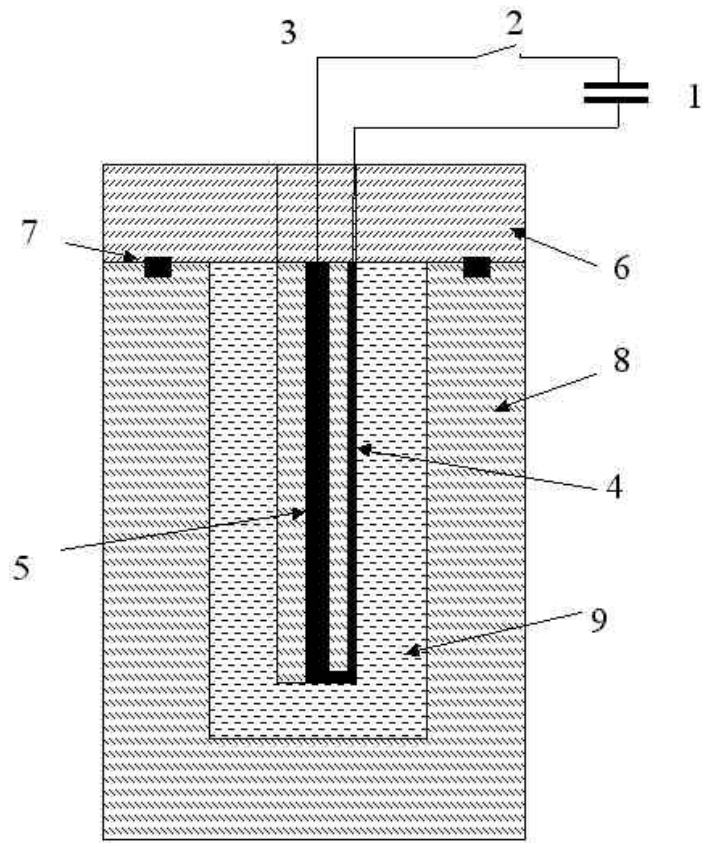

Fig. 1.

1 - capacitor battery, 2 - spark-gap, 3 - cable,

4 -foil, 5 - electrode, 6 - polyethylene cover, 7 - seal,

8 - explosion chamber, 9 - distilled water.

During experimental studies of electric explosion of foils in water, an intensive glowing was found to appear above the dielectric cover. In Fig. 2 we present oscillograms of signals from the photodiode (PD) and the photo-multiplying tube (PMT-35) mounted above the dielectric cover. As seen from the oscillograms, at the moment of the current disruption, which is noted by many authors [3], a glowing emerges above the explosion chamber, which persists over the time period exceeding the current pulse duration by more than 10 times. Since the beginning of the glowing coincides with the voltage drop (see Fig. 2), it is tempting to explain the appearance of the glowing by the ordinary electric break-down in the supplying high-voltage inputs. However, the experimental results described below can not be explained by electric break-down only.

The first argument is that by supplying a static voltage of U~10 kV (just such a tension amplitude appears at the moment of the current disruption), we do not observe electric break-down on the power high-voltage inputs.

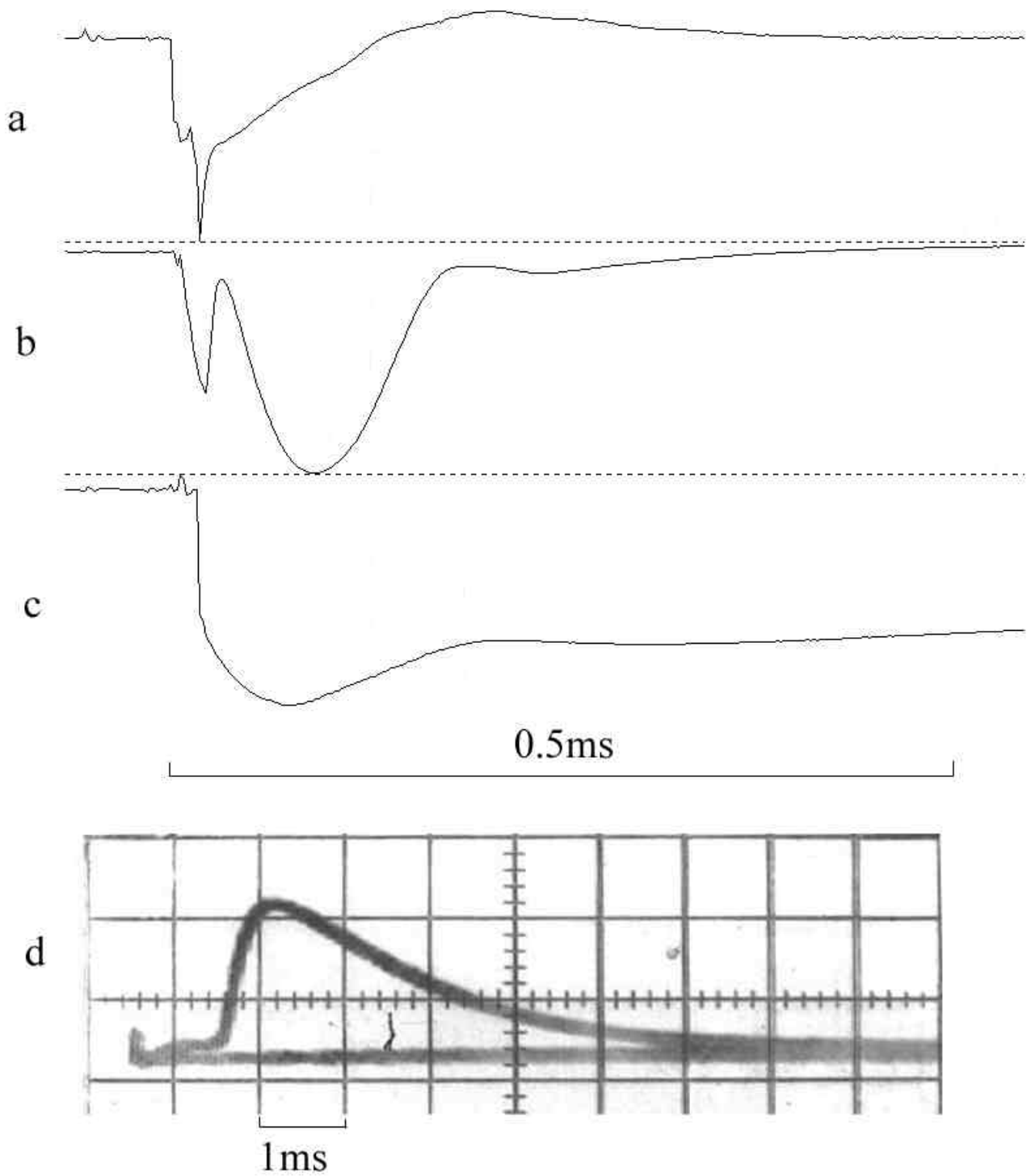

Fig. 2.

a. Tension oscillogram
b. Load current oscillogram
c. Signal from photodiode
d. Signal from PMT-35 attached to interference filter $\lambda=432$ nm

The second argument relates to a significant difference between the duration of the current pulse ~0.15 ms and that of the glowing ~5 ms. However, the recombination time of plasma, which appeared in the air, ~0.1 ms, is much smaller than the observed glowing duration, which does not allow us to explain the observed glowing duration by electric break-down during the current pulse [7]. In the experiments the spectrum of the glowing and the dynamics of the ball-like plasma formation (BPF) were studied.

To study the spectral composition of the radiation, three types of spectrographs were used STE-1 (4300 A-2700 A), ISP-51 (6500 A-4500 A), and DFS-452 (4350 A-2950 A), which allowed us to obtain the time-integrated spectrum (during one shot). The temporal behavior of the narrow spectral fragment was studied with two PMT-35 located 1 meter above the setup with two different interferometric filters ($\lambda_1$=432 nm, $\lambda_2$ =457 nm) (Fig. 3).

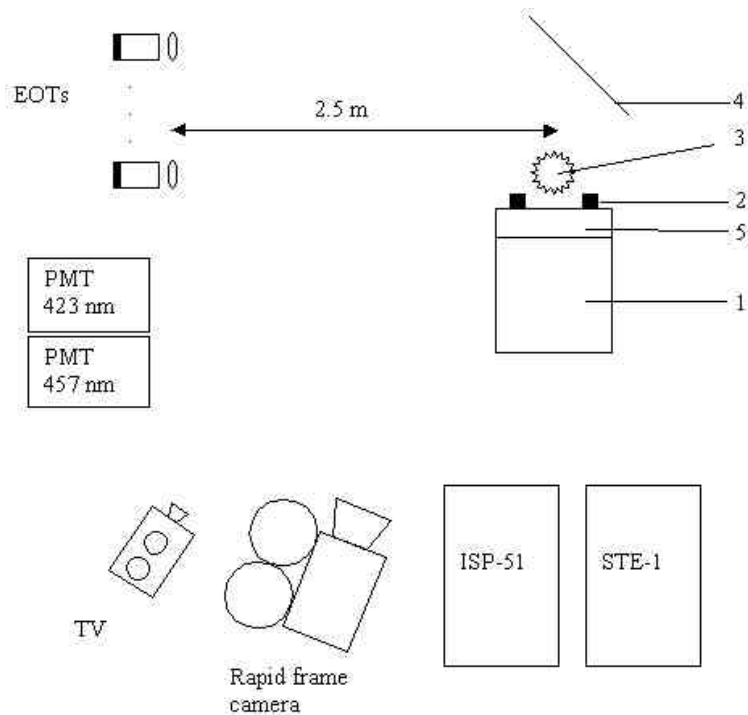

Fig. 3.

Scheme of diagnostics set up

1. Torus
2. High-voltage lead-ins
3. Ball-like plasma formation
4. Mirror
5. Dielectric cover

To register the image of the glowing, three methods with different time resolution were used. The most "quick" method used electron-optical transformers (EOT)[8]. Six EOTs in the frame regime with the exposure time T~130 mcs and the frame delay time T~1 ms allowed to get 6 frames in one shot. The EOTs were set at a distance of 2.5 m from the axis of the setup as shown in Fig. 3. A mirror was mounted ~1 m above the setup by the angle 45 degrees to the vertical line, which permitted us to simultaneously register two projection of the glowing.

To register the glowing an industrial high-speed camera "IMAGE-300" was also used, which enabled us to register 300 frames per second with an exposure time of ~2 ms. For the camera synchronization, a special quartz clocks was designed. The time-integrated wide-field pictures were taken with a standard TV-camera.

Fig. 4b presents an EOT-gram which clearly shows that the glowing appears in the middle between the electrodes above the dielectric cover and has a spherical shape. Using signals from calorimeters, photodiodes, and with account of the results of spectral measurements, the light energy emitted by BPF was estimated to be W~1kJ.

Based on the results of more than 100 tests, the typical dynamics of the spherical-like glowing can be described as follows. At the moment of the current disruption a very bright diffuse glowing emerges in the channel above the setup (Fig. 4a), as if the total space is glowing. Then the glowing fades and in the next spot a spherical-like glowing is clearly seen. No dynamics is observed during the subsequent 3-4 ms (Fig. 4c,d,e), and then the glowing sphere starts dividing into many small "balls". In some experiments the "ball" was noted to rise by 15-30 cm above the dielectric cover and then dissociate (Fig. 4f).

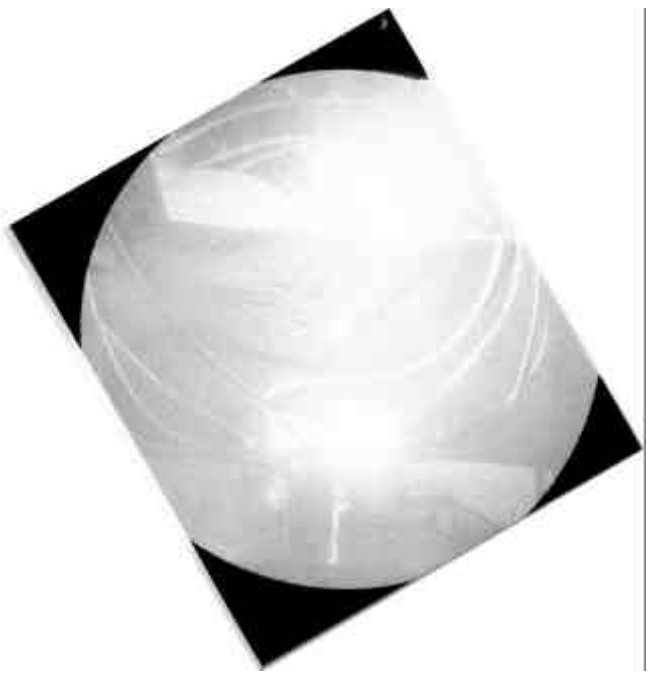

a

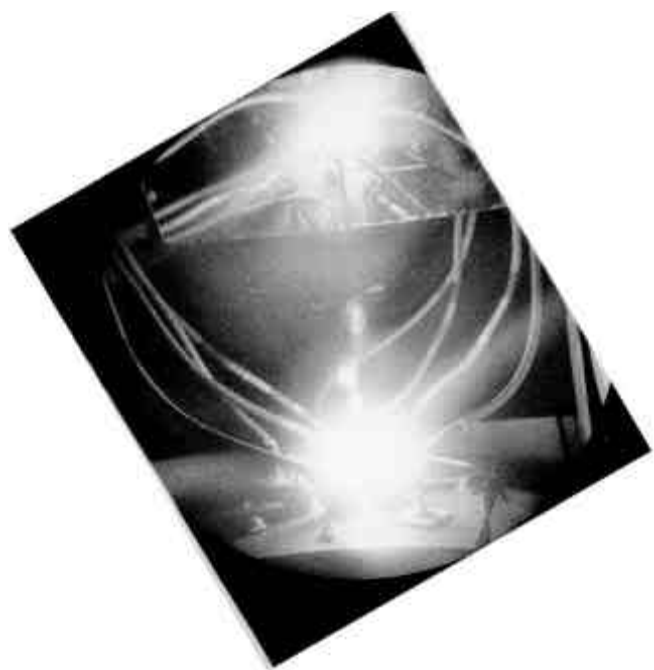

b

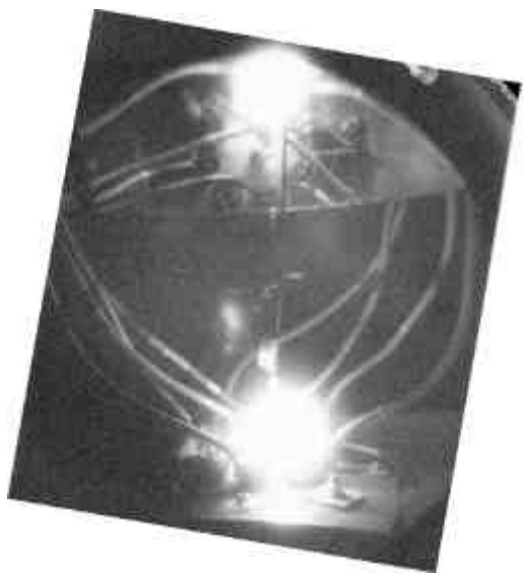

c

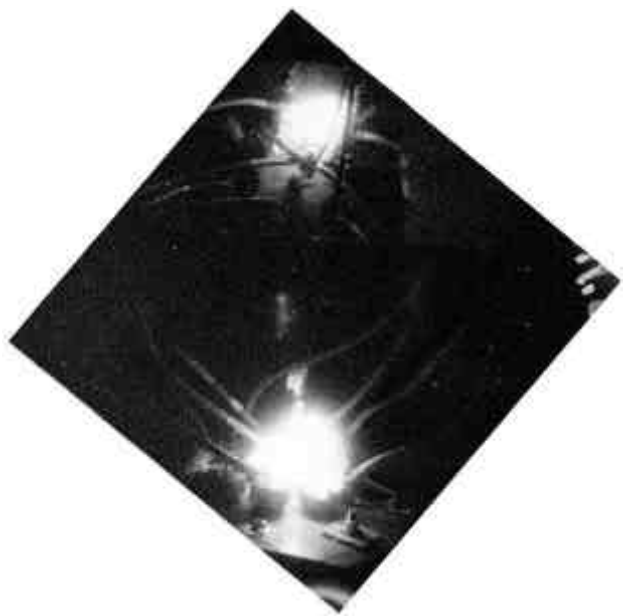

d

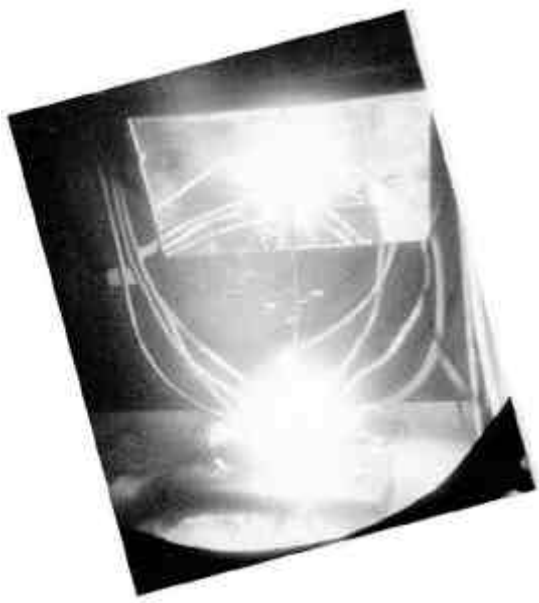 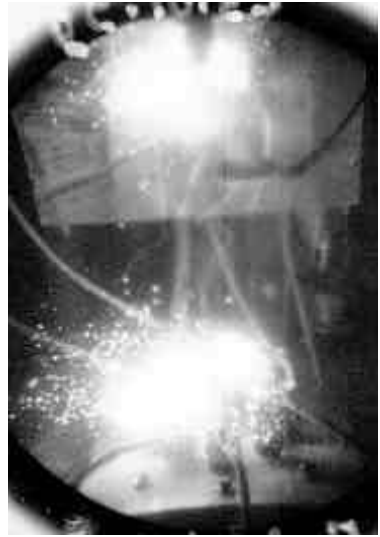

          e                          f

Fig. 4.

Pictures seen on the EOT screens.

Exposure time 130 mcs.

The moment of exposure in Fig, 4a coincides with the time of current pulse.

Time delay between the frames 1 ms

     It should be noted that the characteristic feature of the spherical plasma formation (BPF) is its selectivity with respect to the earth coating of the power and diagnostic cables. In experiments where "earths" of the high-voltage cables were not thoroughly insulated, the BPF often "shorted" on the cable coating, as is seen from EOT-grams. This fact was also supported by measurements of currents, $I$, on a shunt built in the power cable coating. As seen from Fig. 5, at the moment when BPF touches the cable coating, a current emerges in the circuit, the so-called "echo".

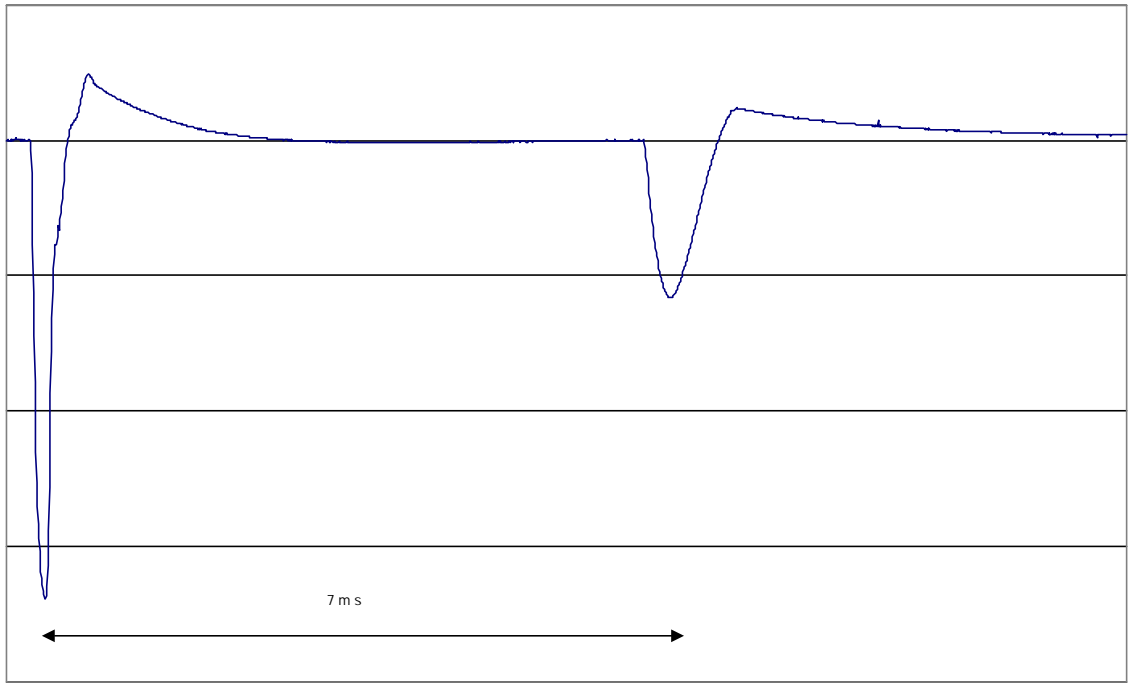

Fig. 5.

The signal form the shunt built in the high-voltage cable coating

Long-lived plasma formations in vacuum were observed in some experiments in various laboratories [9,10]. A distinctive feature of the experiment under discussion is spectral measurements. It is the results of the spectral measurements that became a key to understand the physics of BPF and largely determined the direction of further studies. Fig. 6 demonstrates the fragments of the optical spectra obtained with spectrographs located as shown in Fig. 2. Fig. 6 reveals a spectral line structure in the entire spectral band. In addition, a continuum is also seen, especially in the red part of the optical spectrum.

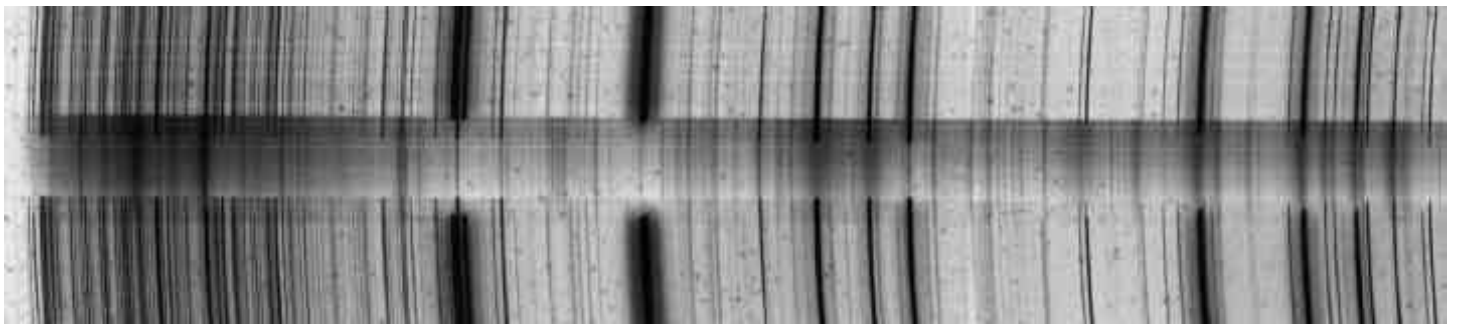

a)

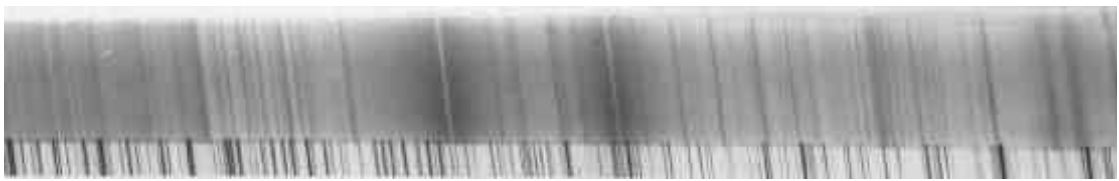

b)

Fig. 6

a). Fragment of the plasma emission spectrum obtained with ISP-51 spectrometer (the upper and bottom spectra are from Cu and Zn standards)
b). Fragment of the plasma emission spectrum (upper) and the iron standard spectrum (bottom) in the wavelength range 3800-4100A obtained with STE-1 spectrometer.

The identification of spectral lines led to two unexpected results. First, no oxygen and nitrogen lines were found (only very weak traces of them were present in separate "shots"), whereas just these lines are always seen during electric discharge in the air. Second, a lot of lines (more than 1000 lines in individual "shots") and, accordingly, a lot of the corresponding chemical elements were discovered. The spectral analysis revealed that most abundant elements in plasma were Ti, Fe (even very weak lines were detected), Cu, Zn, Cr, Ni, Ca, Na. If the presence of Cu and Zn lines in the spectrum can be explained by a sliding discharge along the setup units and power-supply cables, the presence of other lines can not be interpreted. Variation of the experiment conditions, in particular change in mass of the exploding foil, led only to the redistribution in the lines intensity, with the element composition changing insignificantly.

Since titanium foils were exploded, the presence of Ti lines suggested that some fraction of the foil material penetrates through the seals to occur above the setup. In order to check this assumption, the mixture of water and the foil (the "sample" below) was extracted from the channels and subjected to a mass-spectrometer analysis. The results of the analysis are shown in Table 1. The table demonstrates that the original foil consists of 99.7% of titanium. The isotope analysis of the foil shows that Ti isotopes are present in their natural abundance.

| Element | Fraction of atoms, % |
|---------|----------------------|
| Ti | 99.71643 |
| Na | 0.00067 |
| Mg | 0.00068 |
| Al | 0.00921 |
| Si | 0.00363 |
| P | 0.03078 |
| S | 0.03570 |
| Cl | 0.00337 |
| Ka | 0.00253 |
| Ca | 0.03399 |
| V | 0.00195 |
| Cr | 0.00844 |
| Mn | 0.00253 |
| Fe | 0.10613 |
| Ni | 0.04193 |
| Co | 0.00202 |

**Table 1.**

**The original composition of titanium foil.**

The method of studying the "samples" was as follows. The "sample" was first evaporated to yield a dry sediment, which was then carefully mixed to a homogeneous state, and after that was subjected to mass-spectrometer analysis. The mass-spectrometer in use measured atomic masses starting from carbon. Clearly, gases could not be measured with mass-spectrometer. It should be noted that since the mass of the powder under study was about 0.5 grams, it could be visually seen that after the evaporation the "sample" had a inhomogeneous structure.

Unexpected were the results of mass-spectrometer analysis of the "samples", with the typical example presented in Fig. 7a. The total number of all atoms detected in the "sample" is taken for 100%. Fig. 7b shows a histogram of Ti isotope distribution discovered in the same "sample" and in the original foil for comparison (the natural ratio). The total mass of Ti is taken for 100% in histograms in Fig. 7b. Note that the isotope ratio strongly changes in Ti remained after the "shot". The comparison of histograms reveals that the percentage of "shortage" of $Ti^{48}$ in Fig. 7b coincides with the "shortage" of Ti in Fig. 7a.

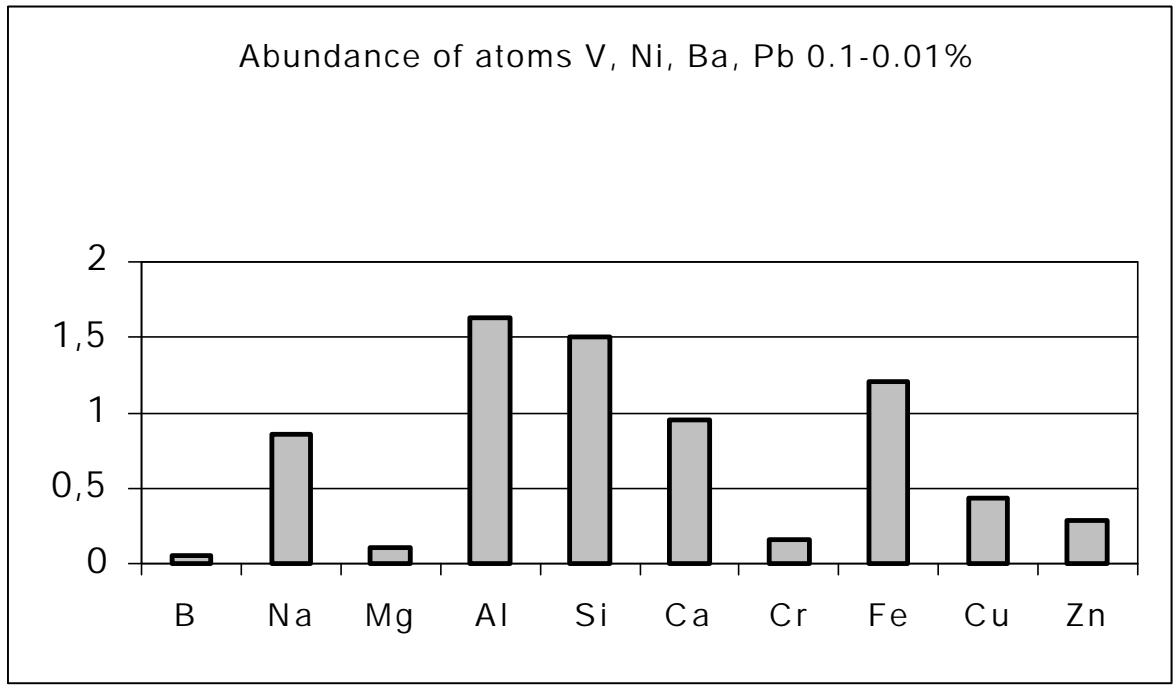

a)

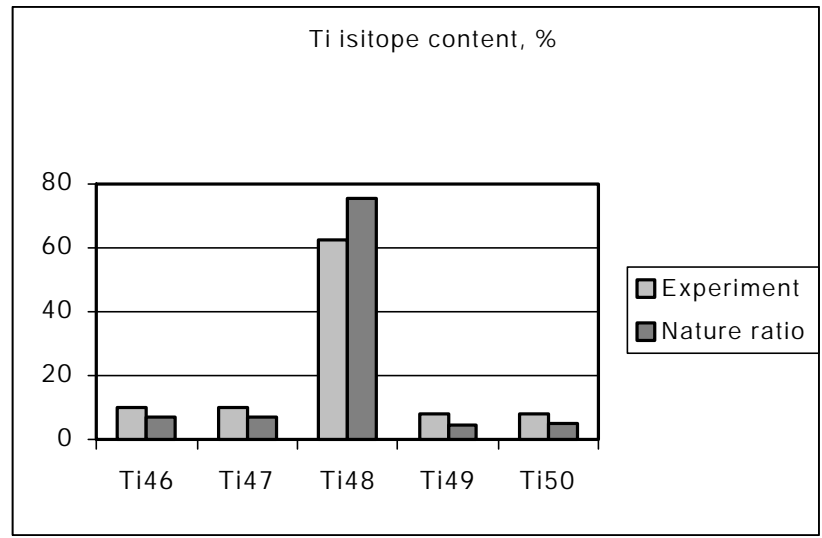

b)

Fig. 7.

Results of mass-spectrometer analysis of products in test 226(Ti load).

a) Percent composition of atoms of "alien" elements in the sample. The fraction of Ti atoms in the experiment products is 92 %.

b) The ratio of Ti isotopes before and after the experiment

All measures were taken to provide the "purity" of the experiment. All electrodes were made of highly purified titanium, new polyethylene caps were used in each "shot". All seals were also made of polyethylene. Since the pressure in the chamber increases during the "shot" due to ohmic heating and chemical reaction of Ti with water, nothing can penetrate the chamber from outside. So only Ti and possibly carbon is expected to be in the "sample". However, in mass-spectra of the "samples" obtained in more than 200 experiments lines of elements ("alien" elements) were detected which were absent in the original material of the exploding foil and electrodes.

To avoid possible errors in measuring mass-spectra, some control "samples" were divided into three parts and directed to three different mass-spectrometers in different laboratories.

Other methods such as electron sounding, X-ray structure, X-ray phase, and X-ray fluorescent analyses, have also been used. The results of electron sounding of a fragment of one of the "samples" are shown in Fig. 8.

Of course, the results obtained by the different methods differ numerically, but qualitatively all methods reveal the presence of a substantial amount of "alien" elements.

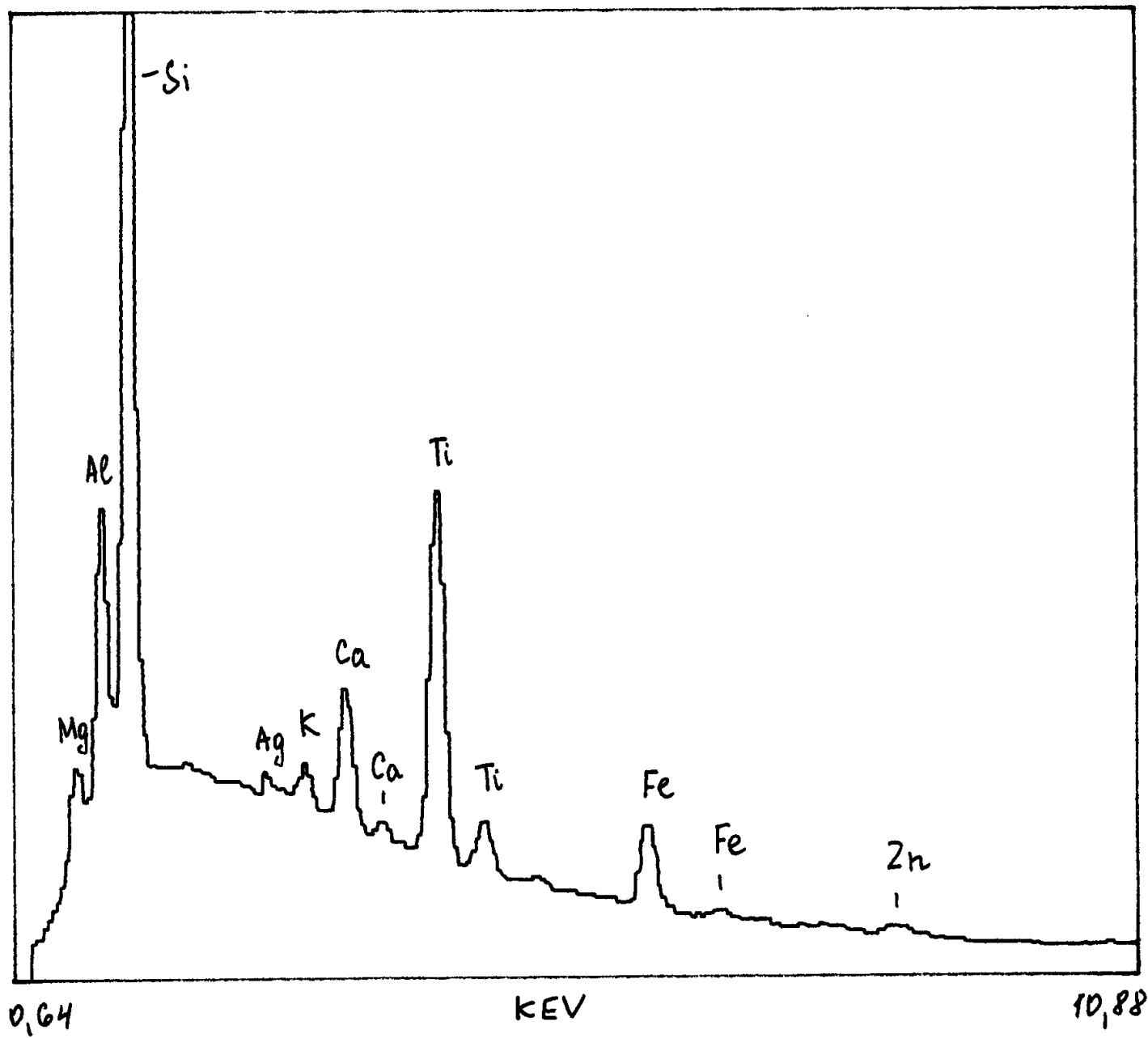

Fig. 8.

The result of electron sounding of a fragment of one of the samples.

The averaged result of mass-spectrometric analyses performed for the "samples" obtained in different shots is shown in Fig. 9. The mean fraction of Ti transformation is 4%. The comparison of histograms in Fig. 7a and Fig. 9 discovers the same elements among the "alien" ones, although their relative contributions in the mass-spectra are of course different. This difference in the specific weight is explained by different conditions of the experiments. The following parameters were changed in the experiments: the energy input into the foil, the number of channels, the mass and size of the foil, an external magnetic field (in some experiments). Thus the experiments using Ti foil as a load revealed the presence of the same "alien" elements. The same conclusion was obtained from spectrometric measurements.

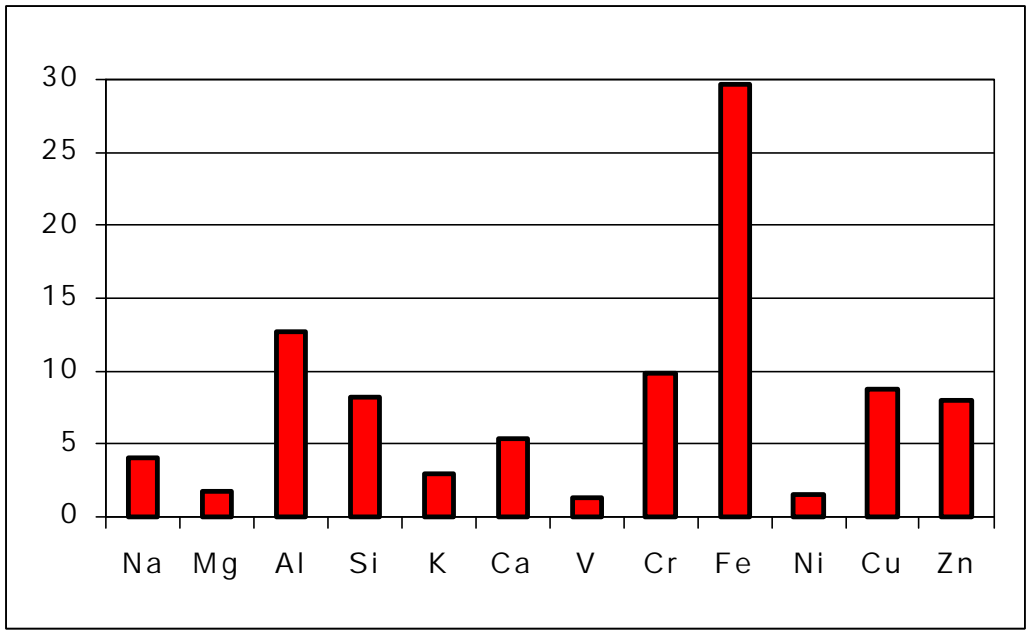

Fig. 9.

The mean percent of "alien" elements atoms in 24 tests (N 169-240) for titanium load.

As noted above, a correlation has been observed between the fraction of mixtures in the "text" and the "skewness" of the isotopic distribution of titanium remaining in the "sample". In all isotopic analyses of sediments the relative fraction of isotopes $Ti^{46}$, $Ti^{47}$, $Ti^{49}$, $Ti^{50}$ was observed to increase and $Ti^{48}$ to decrease. This experimental fact enabled us to suppose that all the decrease of Ti is due to "disappearing" of $Ti^{48}$ isotope. The plot in Fig. 10 is drawn by assuming that the total "disappearing" (burning out) of Ti load is due to only the burning out of $Ti^{48}$. Only experiments with titanium foils as a load were taken to this plot. The plot shows that the points either fall along a straight line y=x, or lie in the upper hemiplane. The last fact indicates that predominantly "alien" elements fly out of the channel, which is in qualitative agreement with spectral measurements from which follows that the fraction of the "alien" elements in plasma is quite significant.

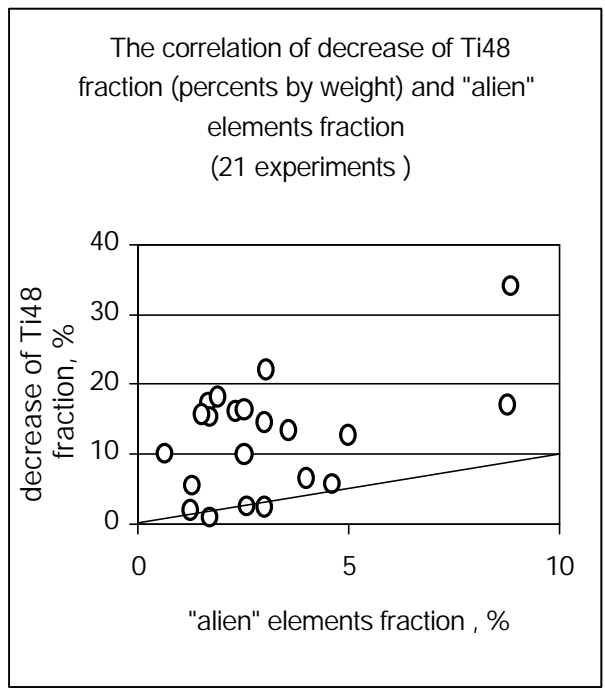

Fig 10.

The correlation of decrease of Ti48 fraction and increase of the "alien" elements fraction (percents by weight).

Fig. 11 presents the histogram of the mean composition of the products for experiments with zirconium Zr foil as a load. The original zirconium foil contained 1.1% of niobium, which was subtracted from the final product composition. Comparing Fig. 9 and Fig. 11 suggests each original loads produces individual spectrum of chemical elements. This statement holds for other foils (Fe, Ni, Pb, V, Ta), which were used in other experiments.

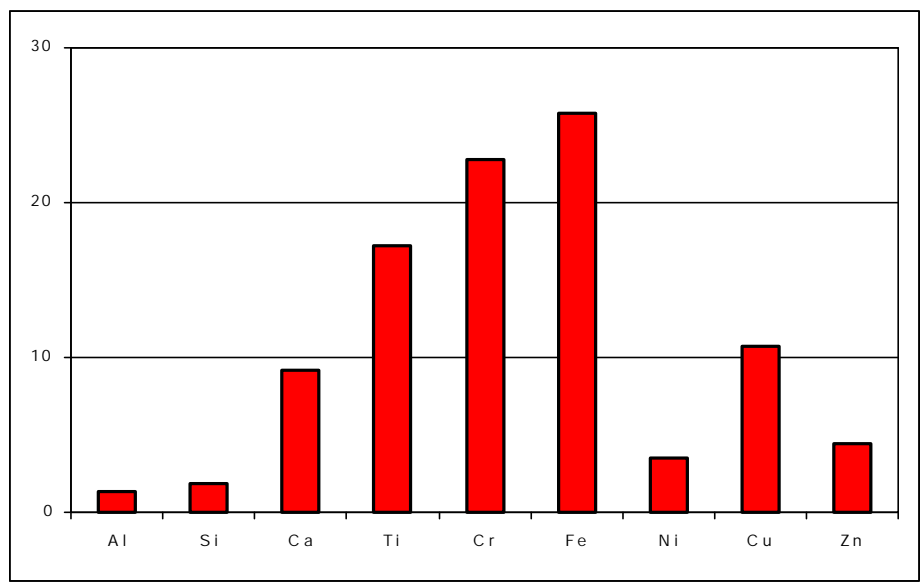

Fig. 11.

The mean percent of atoms of "alien" elements in 5 tests with zirconium load.

Since the transformation of elements must have been associated with some radioactive emission, intensive search for gamma-radiation and neutrons have been done. To register gamma-radiation integral dosimeters, X-ray films, and CsI scintillator detectors with PMT-30 have been used. No significant X-ray flux has been detected in any experiment. As followed from mass-spectrometric results, the number of acts of transformation was $10^{19}$-$10^{20}$ per shot, so clearly even one gamma-ray photon per transformation act would led to an enormous gamma-ray flux of P~$10^{20}$.

In order to register neutrons we used 2 plastic scintillator detectors with PMT-30. The detectors were mounted at a distance $\alpha_1$~0,4m, $\alpha_2$~0,8m from the setup axis. In Fig. 12 the typical signal from PMT-30 is presented with a duration of T~100 ns. Such a short duration was a great surprise since the current pulse duration was ~ 20 ms. In order to measure the time of arrival of particles, a special transformer was designed which formed a standard pulse of ~10 ms from external signal with ~10 ns. Thus the studied pulse from the detector triggered the oscillograph, then was directed to the oscillograph input through a delay line, and only after that entered the transformer and ACP. The time delay in the turn-on of two oscillographs registering signals from two plastic detectors allowed us to measure the radiation propagation velocity. It was found to be V~20-40 m/s. Such a low velocity precluded signals to be neutrons, since then they must have been ultra cold and could not reach the detector and moreover to overcome the light-shielding cover made of aluminum. To understand the nature of the radiation and obtain its "self-portrait", a method using photo emulsions was applied.

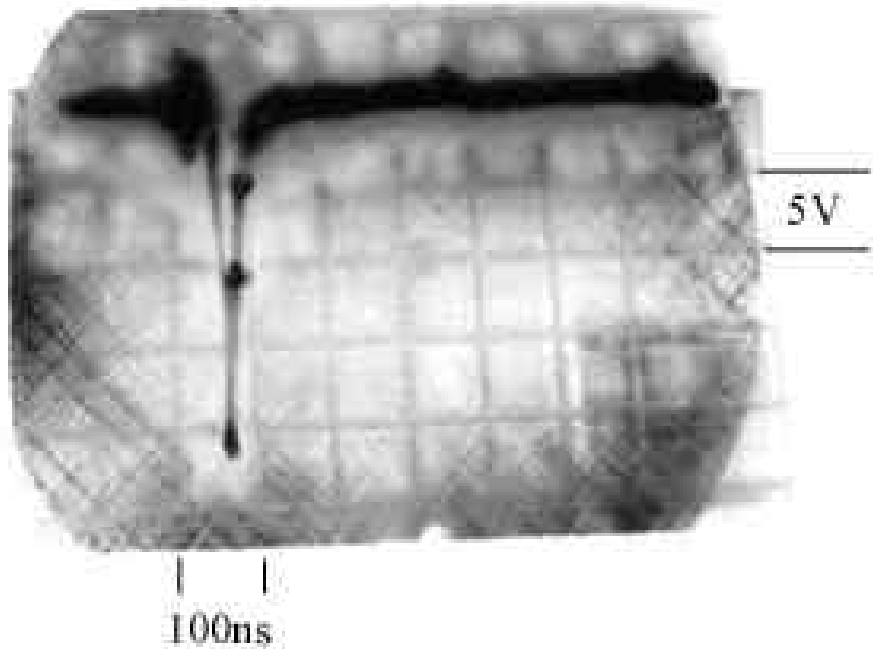

Fig. 12.

The signal from polystyrol scintillator detector.

The following materials have been used in the experiment: a fluorographic film RF-ZMP with sensitivity 1100 $R^{-1}$ at a level 0.85 above the haze, a radiographic medical film RM-1MD with sensitivity 850$R^{-1}$ at a level 0.85 above the haze, nuclear photo plates of type R with a thickness of the emulsion layer of 100 mcm, high-resolution photo emulsions with a sensitivity ~0.1 GOST units and a resolution of up to 3000 lines/mm.

All the materials were processed after the exposure in the corresponding developers: the fluorographic films in D-19 developer during 6 min at temperature 200C, the plates in a phenydon-hydrochinon developer using isothermal method.

The inspection of the processed materials revealed micro and macro effects. Macro effects included those that can be seen by naked eye or using a magnification glass with up to 5 times magnification. Micro effects included those seen under magnification from 75 to 2025 times. Films and photo plates were set at different distances from the center of electric explosion (from 20 cm to 4 m) in radial and normal planes assuming cylindric symmetry of the experiment (Fig. 13).

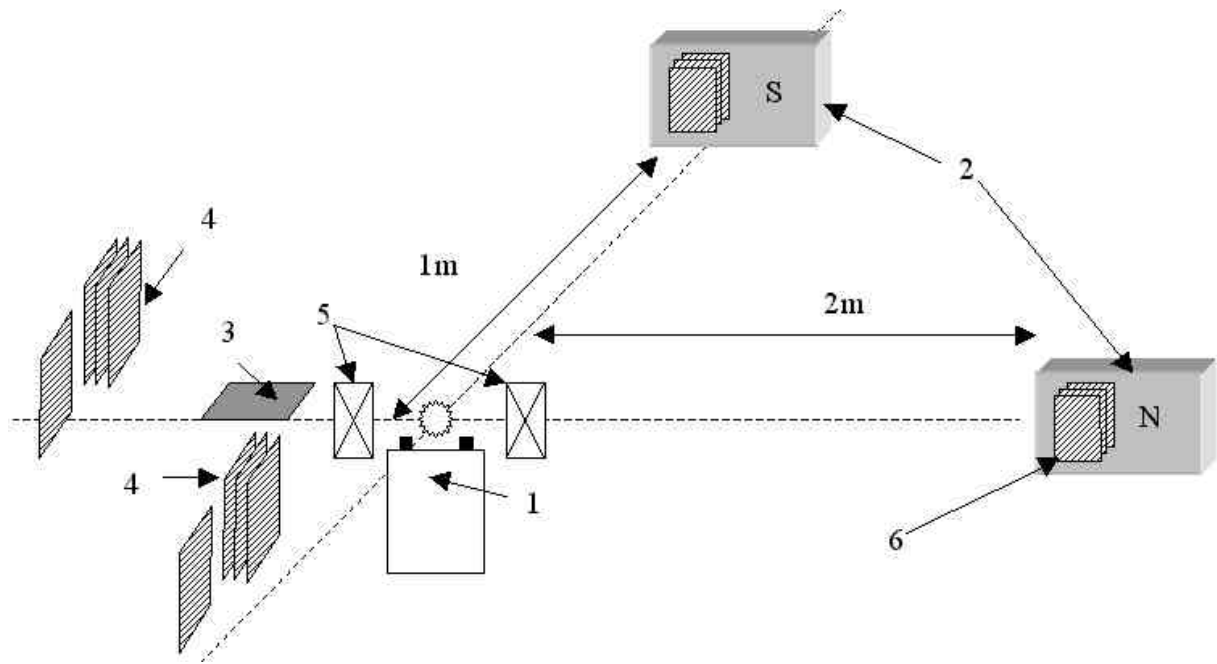

Fig. 13. The scheme of photo detectors location

1 - the site of electric explosion of foils

2 - permanent magnets

3 - the plate with nuclear emulsion

4 - films

5 - magnetic field coil

6 - films near the permanent magnet

All the materials were carefully wrapped in the black paper, which had preliminary been inspected to have damages. After the exposure in the setup and developing photodetectors, the paper has been expected once again.

The very first experiments revealed a wide variety of track forms, including continuous straight lines, dumbbell-like ("caterpillar") tracks, and long tracks with a complex form similar to spirals and gratings.

Fig. 14a demonstrates a typical very long (1-3 mm) track similar to that of a caterpillar or a tire-cover protector. These tracks are characterized by having the second parallel trace with darkening and length different from the main one. The track presented in Fig. 14a was formed on the fluorographic film RF-ZMP with emulsion layer thickness 10 mcm. In Fig. 14b a magnified fragment of the track is shown which clearly demonstrates a complicated pattern. Notably that with a grain size of ~1 mcm, the track width is about ~20 mcm. The estimate of the energy of particles obtained from the darkening area is E~700 MeV assuming Coulomb braking. Taking into account the position of the photo detector (shown in Fig. 13) and the track size, the track can not be explained by alpha, beta, or gamma-radiation (recall that RF-film is wrapped in the black paper and is surrounded by the air). To check the nature of the "strange" radiation, the remnants of the foil and water were extracted from the channel after the explosion and put into a Petri cap (the "sample"), and the photo detector was set 10 cm away as shown in Fig. 15a. The film RF pressed to fiber glass washer was used, and the entire detector was enveloped in the black paper. The fiber glass washer was used because in the previous experiments we noted that the "strange" radiation clearly demonstrates properties of the transition radiation. The exposure time was T~18 hours. The result is presented in Fig. 15b. The inspection of Fig. 14 and 15 suggests that the darkening of the film in both cases was due to identical reasons. This in turn implies that the radiation was not caused by acceleration and had a nuclear origin. It should be noted that the position of the detector planes normal to the radius vector in both cases allows the interpretation that the source of the registered emission moved with a non-zero angular velocity.

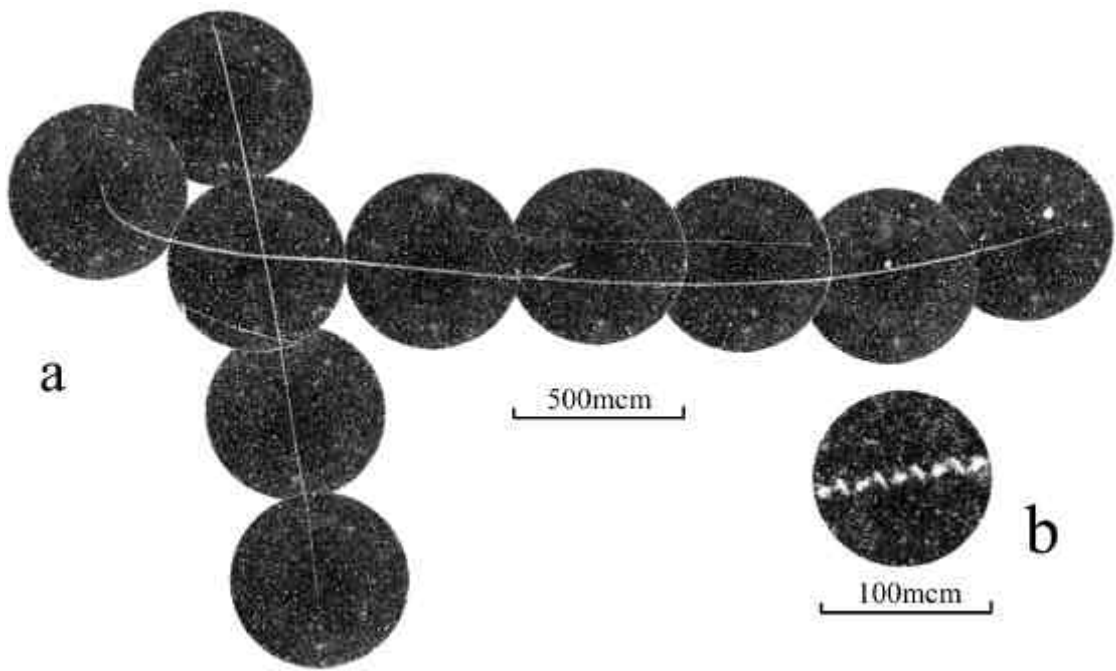

Fig. 14.

The typical track on the film.

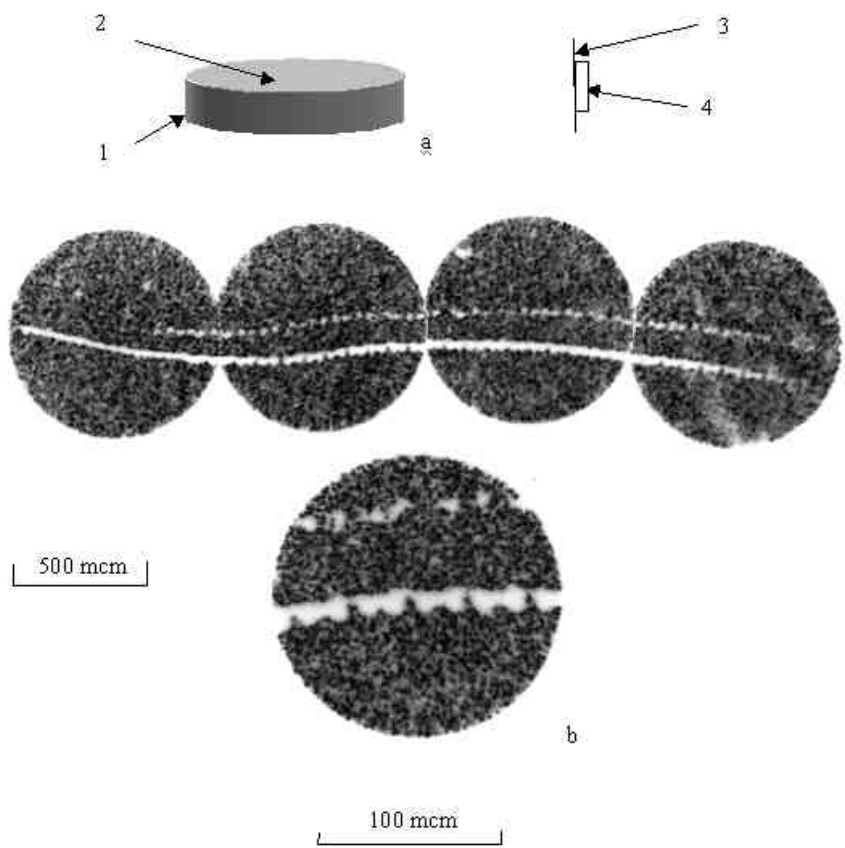

Fig. 15.

a) The scheme of experiment. 1 - Petri cap; 2 - the sample; 3 - film; 4 - fiber glass.

b) The track and its magnified fragment.

The detection of the same tracks using nuclear emulsions with thickness 100 mcm permits us to state that the source caused the darkening flies strictly in the photo emulsion plane, since the depth of the beginning and the end of the track inside the emulsion differs by less than 10-15 mcm.

Assuming the electric pulse in Fig. 12 and the track are due to one and the same reason and accounting for the track length and the pulse duration, we arrive at the estimate of the radiation source velocity $\beta=10^{-3}$.

A series of experiments was performed to study the effect of the external magnetic field on the picture observed. Using a magnetic coil located as shown in Fig. 13, a weak magnetic field H~20 G was imposed in the site of explosion.

The photodetectors were set as shown in Fig. 13(3). The typical tracks registered are shown in Fig. 16(a,b), with a nuclear photo emulsion as photo detector. It is seen from the Figure that the track strongly changes and the trace becomes "comet"-like in shape.

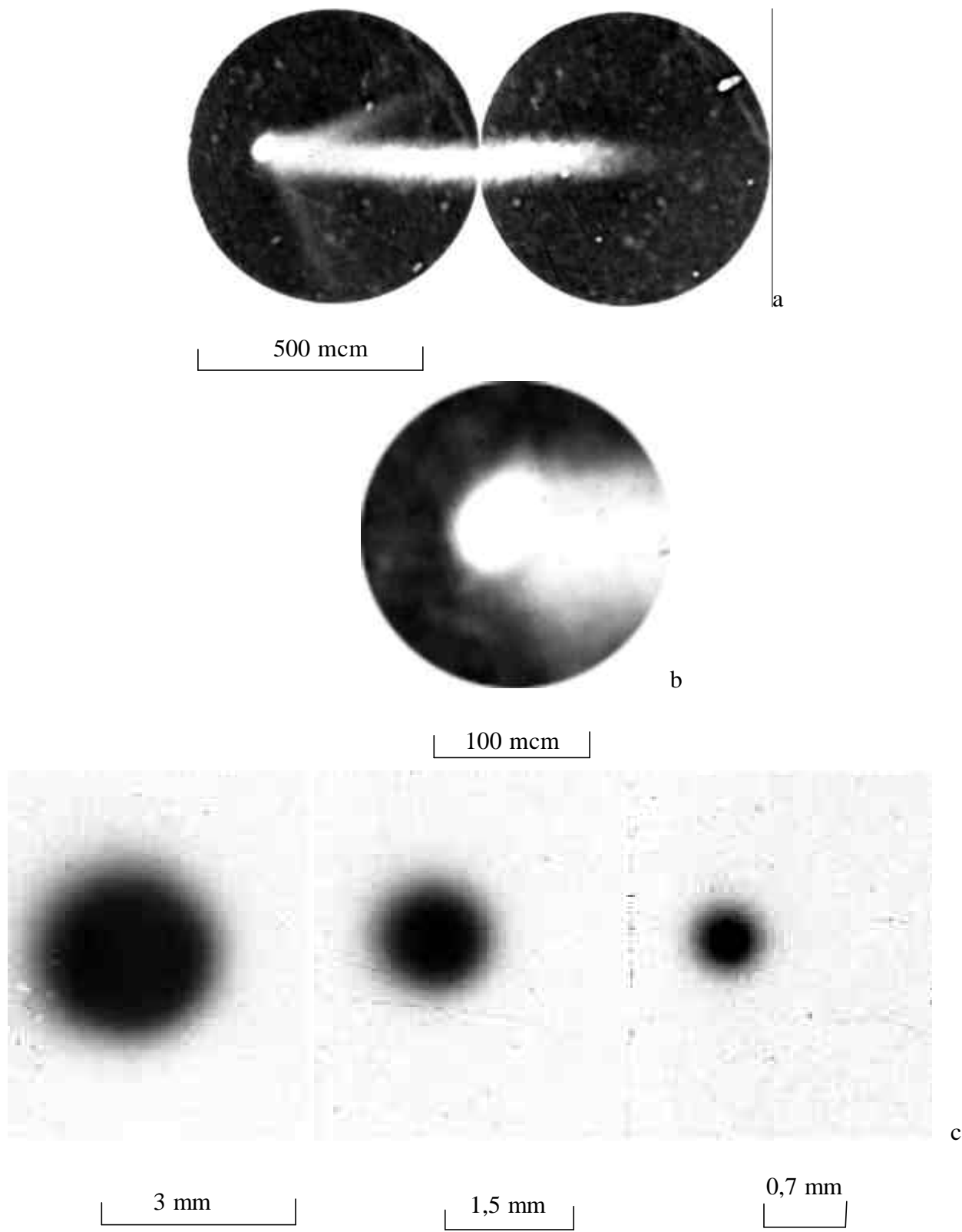

Fig. 16.

a) The "comet"-like trace

b) A magnified fragment of the "comet head"

c) The "flaky" trace

A more detailed study of the structure under microscope with 225 times magnification allowed us to single out a round head (Fig. 16b) with the darkening density D>3 and a long tail with decreasing darkening density similar to the "comet tail" (Fig. 16a). Six such "comets" were detected inside the area 4 cm$^2$. Their sizes varied from 300 mcm to 1300 mcm and the energy of particles derived from the darkening area reached E~1GeV.

In some experiments the detector (and not the entire setup) was put in the magnetic field. Fig. 16c shows the traces obtained by a photo detector consisting of three RF-films folded together and located near a samarium-cobalt magnet (B ~ 1.2 kG) as shown in Fig. 13(b).

The positions of darkenings in Fig. 16c coincide geometrically, which exclude them to be artifacts. The energy absorbed in the three films with account only the photo emulsion layer thickness (~10 mcm) is estimated to be E~700 MeV.

Thus it is clearly seen that the magnetic field affects the "strange" radiation. In addition to the tracks shown, we have registered some tracks quite different in shape from the "classical" ones. Part of these tracks not presented in this paper is very similar to scratches or ink spots. Exactly the same tracks were observed in experiments by Matsumoto [11] using other types of photo detectors and in experiments with breakdown in water. This precludes us from unconditionally relating them to artifacts. However, already today we can say with a significant reliability that one or two types of particles can hardly explain all tracks which appear during the registration of the "strange" radiation.

**Discussion of experimental results. A magneto-nucleon catalysis hypothesis.**

Let us try to emphasize the main properties of the observed phenomena. Apparently, it is the symmetry of location of plasma channels that was the reason for the BPF appearance, which in turn allowed the dynamics and optical spectrum of the emergent radiation to be studied. >From Fig. 5 a purely visual association between the BPF and ball lightning arises. There are at least two major problems in physics of ball lightning: the reason for appearance and failure to explain the source of radiation. The experimental results reported here can suggest a reason for the BPF appearance. Indeed, the optical spectra analysis and mass spectrometry results are in qualitative agreement. This allows us to suppose that during the process a fraction of materials leaks through the seals. However, even if not to analyze how plasma penetrates through the seals, two questions remains. Why all the plasma coalesces into a ball and does not dissociate and why this occurs in such a short time since we do observe the BPF already on the first EOT frame (in some "shots").

One can try to explain the BPF using the cluster model [9] of the fractal ball model [12], however we have looked for a hypothesis which can explain all experimental facts. Such a hypothesis, in our opinion, could be the formation of magnetically-charged particles (magnetic monopoles). The first attempt to explain the ball lightning by magnetic monopoles was done in paper [13]. This paper suggested to explain the properties of ball lightning by Rubakov's effect [14] predicted for super-heavy monopoles which should exist in the Great Unification Theory [15,16] (the so-called GUT monopoles). In our opinion, the experimental results obtained in our paper provides no serious support for the hypothesis [13]. However the assumption that magnetic monopoles form in the plasma discharge in water could be one possible reason for the obtained experimental results, how unusual this assumption might appear.

Indeed, a wide track similar to the trace of a "crawling caterpillar" [17] was expected just for classical monopoles [18,19]. The estimate of energy dissipated in the emulsion by radiation E~1 GeV coincides with that expected for magnetic monopoles. A visual change in the track shape observed in the presence of magnetic field also gives support to this assumption.

To confirm directly the fact of magnetic monopole creation in plasma discharge an experiment was performed using the idea from paper [20], in which iron foils were suggested to be taken as a monopole trap. In our experiment we used three $^{57}$Fe foils, which has an ideal structure and a significant field near the nucleus.

Since both N and S magnetic monopoles must appear, the foils under study were located near different poles of a strong magnet with the magnetic field strength H~1 kG in anticipation of selection of monopoles. Thus N-monopoles were expected to attract by the S-pole and S-monopoles by the N-pole of the magnet. The magnets were set at a distance of h~70 cm away from the site of electric explosion. The third foils was used as a standard.

Due to a large magnetic charge the monopoles "captured" in the trap must change the magnetic field near $^{57}$Fe nucleus, which can be measured by Moessbauer effect for a sufficiently large number of the "trapped" monopoles.

The results of measurements showed that in the foils located near the N-pole the absolute value of the superfine magnetic field increased by 0.24 kG. In another foil (S) it decreased by approximately the same amount 0.29 kG. The measurement error was 0.012 kG.

Fe-standard:   H = 330,42 kG

Fe-north -N : H = 330,66 kG, $\Delta_N$ = 0,24kG

Fe-south -S :  H = 330,13kG, $\Delta_S$ = -0,29kG

Taking into account that the magnetic field in $^{57}$Fe has the opposite sign with respect to its magnetization, we can state with certainty that S-particles (at the N-pole of the magnet) increase the negative superfine field, while the particles with the opposite sign decrease it, with the relative change being $\sim 8 \cdot 10^{-4}$.

From analysis of Moessbauer spectra of ferromagnetics, the absorption line width is known to increase. This phenomenon is related to inhomogeneity of internal magnetic fields near nuclei. An analysis of spectra of irradiated foils revealed an additional absorption line broadening comparable with the ordinary magnetic broadening. This possibly relates to a chaotic absorption of the monopoles in the lattice of iron.

Fe-standard:    r_1=0.334/0.300/0.235 mm/s

Fe-north -N   :  r_1=0.363/0.328/0.250 mm/s

Fe-south -S   :  r_1=0.366/0.327/0.248 mm/s

The measurements error 0.003 mm/s.

No quadruple shift of levels is discovered, i.e. no change of the electric field gradient in crystal is observed. The results of this experiment strongly support the magnetic monopoles hypothesis. Unfortunately, these measurements do not allow to decide whether magnetic monopoles have electric charge.

Using the hypothesis of magnetic monopole formation we can suggest that the observed BPF are magnetic clusters. In analogy with [9] we can suppose that the role of ion is played by a monopole coupled with the foil atom nucleus, and the solvation occurs due to interaction of the monopole magnetic charge with magnetic moment of oxygen atoms.

The main regularities experimentally observed during the transformation of chemical elements can be summarized as follows.

1. The transformation occurs predominantly with even-even isotope, which leads to a notable distortion of the original isotope content.

2. Experiments with foils made of different chemical elements have shown that they transform into individual spectra of elements, and the statistical weight of each element is determined by concrete conditions.

3. For the set of chemical elements resulted from the transformation there is a minimal difference $\Delta \text{Å}_b$ between the binding energy of the original element and the mean over spectrum binding energy of the formed elements. The difference of binding energies $\Delta E_b = E_{orig} - E_{form}$ (with account of the real isotope ratios) calculated from mass-spectrometric measurements in different tests, falls within the range $\Delta E_b < 0.1$ MeV/atom, which is clearly due to mass-spectrometric measurements errors.

4. No increase in the binding energy difference $\Delta E_b$ as a function of the transformation fraction of the original chemical element has been detected.

5. All nuclei of chemical elements resulted from the transformation are in the ground (non-excited) state, i.e. no appreciable radioactivity has been found.

To explain the element transformation, we have put forward the working hypothesis of magneto-nucleon catalysis (MNC). We introduced this term to designate the process which supposedly occurs in the plasma channel. The essence of MNC is that the magnetic monopole with a large magnetic charge and an even small kinetic energy can overcome Coulomb barrier and become bound with atomic nucleus. The MNC must have many common features with muon catalysis [21], in which the Coulomb barrier is substantially decreased due to the large mass of mu-meson. Since the magnetic monopole seems to be a stable particle, the MNC may be more effective.

The experiments established that the transformation and hence the MNC occurs only inside the plasma channel.

In conclusion, the authors greatly acknowledge the members of the staff of "RECOM" A.G. Volkovich, S.V. Smirnov, V.L. Shevchenko, S.B. Shcherbak, and to members of the staff of "Kurchatov Institute" V.A. Kalenskii, R.V. Ryabova, Yu.P. Dontsov, B.V. Novoselov, and A.Yu. Shashkov for assistance in the experiments. We also deeply thank A.I. Voikov for financial support of the present study and A.A. Rukhadze for support and help.